%% file: evol-v7.tex
\documentclass[useAMS,usenatbib]{mn2e}

\usepackage{graphicx}  %need this for the images
\usepackage{tabularx}
\usepackage{xcolor}
\usepackage{soul}
\def\note #1]{{\bf #1]}}
\newcommand{\apj}{ApJ}
\newcommand{\apjl}{ApJL}
\newcommand{\aap}{A\&A}
\newcommand{\aj}{AJ}

\newcommand{\mnras}{MNRAS}

\newcommand{\nat}{Nature}
\newcommand{\apjs}{ApJS}
\newcommand{\nuMax}{$\nu_{\rm max}$}
\newcommand{\DeltaNu}{$\Delta\nu$}
\newcommand{\eps}{$\epsilon$}
\newcommand{\UP}{Universal Pattern}
\newcommand{\LL}{$12\,\mu \rm Hz$} % below this limit is RGB
\newcommand{\UL}{$125\,\mu \rm Hz$} % above this limit is RGB
\newcommand{\SC}{$50\,\mu \rm Hz$} % secondary clump divide frequency
\newcommand{\PL}{75\,s} % secondary clump divide period
\newcommand{\IL}{\textit{inter-low}} % interlow
\newcommand{\IH}{\textit{inter-high}} % interhigh

\newcommand{\uHz}[1]{ $#1\,\mu \rm Hz$}
\newcommand{\ch}[1]{{#1}}  % corrections in bold for the referee
\newcommand{\OPS}{$\Delta P_{\rm obs}$}
\title[Evolutionary State of Red-Giant Stars]{A New Method for the Asteroseismic Determination of the Evolutionary State of Red-Giant Stars}
\author[Yvonne Elsworth,  Saskia Hekker,  Sarbani Basu, \& Guy Davies]{%
Yvonne Elsworth$^{1,2,}$\thanks{E-mail:y.p.elsworth@bham.ac.uk},
Saskia Hekker$^{3,2}$,
Sarbani Basu$^{4}$  
\& Guy Davies$^{1,2,}$\\
$^{1}$School of Physics and Astronomy, University of Birmingham, Birmingham B15 2TT, UK\\
$^{2}$Stellar Astrophysics Centre, Department of Physics and Astronomy, Aarhus University, Ny Munkegade 120, DK-8000 Aarhus C,
Denmark\\
$^{3}$Max-Planck-Institut f{\"u}r Sonnensystemforschung, Justus-von-Liebig-Weg 3, 37077 G{\"o}ttingen, Germany\\
$^{4}$Department of Astronomy, Yale University, New Haven, CT
06511, USA\\
}

\begin{document}

\date{ submitted 28th September 2016}

%\pagerange{\pageref{firstpage}--\pageref{lastpage}} \pubyear{2002}

\maketitle

\label{firstpage}

\begin{abstract}
Determining the ages of red-giant stars  is a key problem in stellar astrophysics.
One of the difficulties in this determination is to know the evolutionary state of the individual stars -- i.e. have they started to burn Helium in their cores? That is the topic of this paper.
Asteroseismic data provide a route to achieving this information.
What we present here is an highly autonomous  way of determining the evolutionary state from an analysis of the power spectrum of the light curve. The method is fast and efficient and can provide results for a large number of stars. 
It uses the structure of the dipole-mode oscillations, which have a  mixed character in red-giant stars, to determine some measures that are used in the categorisation. 
It does not require that all the individual components of any given mode be separately characterised.  
Some 6604 red giant stars have been classified. Of these 3566 are determined to be on the red-giant branch, 2077 are red-clump and 439 are secondary-clump stars. We do not specifically identify the low metallicity, horizontal-branch stars.
The difference between  red-clump and secondary-clump stars is dependent on the manner in which Helium burning is first initiated. 
We discuss that the way the boundary between these classifications  is set may lead to mis-categorisation in a small number of stars. 
The remaining 522 stars were not classified either because they lacked sufficient power in the dipole modes  (so-called depressed dipole modes) or because of conflicting values in the parameters.
\end{abstract}


\section{Introduction}
In a classical Herzsprung-Russell diagram (HRD) stars burning hydrogen in a shell around an inert Helium core  and stars with Helium-core and hydrogen-shell burning   occupy overlapping parameter spaces. This is due to the fact that these stars can have similar surface characteristics such as effective temperature, surface gravity and luminosity but  different internal conditions. 
Hence, from `classical' observations such as brightness and temperature it is often not possible to distinguish between red-giant-branch stars of about 10-12 R$_{\odot}$ and core-Helium-burning stars of a similar radius. It is this problem that we address in this paper.

In this paper we describe and use a new method to identify the evolutionary state of red-giant stars.  We are concerned with stars that are either ascending the red-giant branch (RGB) or stars which have started to burn Helium in their cores,  but have not evolved to later stages. We do not identify asymptotic-branch (AGB) stars as such and will probably, by default, classify them as RGB stars.
There are two categories of core-Helium burning stars  considered here --- red clump (RC), these are stars that \ch{had}  degenerate cores at the tip of the giant branch and \ch{underwent}  Helium flash, and secondary-clump (SC) stars for which He burning \ch{began} in a quiescent manner.
For the sake of simplicity we introduce a  notation that allows us to specify all  stars that are burning Helium in their cores be they red-clump, low-metallicity horizontal-branch or more massive, secondary-clump stars that did not pass through the Helium flash with a single acronym. We designate all these as HeB stars.

\subsection{Asteroseismology as a Tool}
\label{framework}
Asteroseismology -- the study of the internal structures of stars through their global oscillations is a well-developed technique.
In addition to main-sequence and sub-giant stars, it has been successfully applied to RGB and HeB stars \citep[e.g.][]{deridder2009, bedding2010, huber2010, hekker2011, corsaro2012, hekker2013,mosser2014, deheuvels2015,stello2016}. 
The power spectrum formed from timeseries data contains peaks which are evidence of normal modes of oscillation in the star. The modes are characterised by an order ($n$) dependent on the radial behaviour,  a degree ($\ell$) and an azimuthal order ($m$). 
The degree and azimuthal order together describe the orientation of the nodal lines on the surface of the star.
Modes with a degree of 1 (i.e. $\ell=1$) are known as dipole modes.

The peak value of the oscillations in the power spectrum is located at a frequency known as \nuMax, and different  orders in the spectrum with the same degree are separated by \ch{approximately} \DeltaNu. 
The radial modes are acoustic modes (p modes) and in RGB and HeB stars, all higher degree modes have a mixed character where pressure and buoyancy contribute to the restoring force. 
In what follows we concentrate on the features in the spectra of red-giant-stars.

\ch{The power spectrum can be described by the formulation introduced by 
Tassoul \citep{tassoul1980},  in which the positions of peaks of a given order and degree are predicted by a  formula that includes the  parameter, \eps , which is a constant that describes the  shift in the position of the spectrum in frequency.
Here we use a development of this, the so-called \UP\ \citep{mosser2011up}, where all the parameters in the formulation are functions of \DeltaNu\  .}

As mentioned above, the $\ell=1$ modes in giant stars are mixed modes.
These carry information from both the core
(gravity modes) and the stellar outer regions (acoustic modes).
\ch{\citet{hekker2009} alluded to the presence these modes in their analysis of data from the 
\textit{CoRoT} satellite. 
\citet{bedding2010} reported the first detection of mixed modes in red giants.  
Subsequently, \citet{beck2011} measured period spacings and then 
\citet{bedding2011} used the typical spacing in period between mixed modes
to distinguish between RGB and HeB stars.
(see also \citet{mosser2011mm}).}
HeB stars have a larger  spacing in period compared to that in RGB stars.
These differences are partly caused by changes in the density differences between the core and the outer regions \citep{montalban2010}, as well as due to the fact that in HeB stars the core is (at least partly) convective \citep{jcd2014book}. 

We note here that there is a significant difference between the  asymptotic period spacing of a pure g mode and the spacing observed in a typical power spectrum. The observed period spacing is always smaller than the asymptotic one due to the coupling of the pure gravity modes with the pure acoustic mode. This coupling weakens with increasing frequency separation between the gravity mode and acoustic mode with which it couples, and hence the observed period spacing increases towards the asymptotic value for gravity dominated modes.
In this paper we are concerned with an observed period spacing that is an average over the \ch{observed} mixed modes and not with the asymptotic value. \ch{As shown in the first papers on the observation of the period spacing of mixed modes and referred to earlier, characterisation} based on the observed period spacing provides sufficient information to determine the evolutionary state of the stars, which is the aim of this work.

\subsection{Existing methods to determine evolutionary state}
Currently, different methods exist to determine the evolutionary state of red-giant stars.  
They rely on various features in the  power spectrum.
The earliest method used in the determination of evolutionary state is that by
 \citet{bedding2011}. 
\ch{They determined the most prominent period spacing by taking the power spectrum of the power spectrum, where the latter was expressed in period rather than frequency and set to zero in regions not containing the $\ell=1$ modes.}
\citet{mosser2011mm} also concentrated on the dipole modes and they applied the  envelope autocorrelation function (EACF) \citep{mosser2009} with a very narrow filter to isolate the dipole modes.

A different approach was taken by \citet{kallinger2012} which concentrated on the $\ell=0$, radial modes. They were able to demonstrate that a determination of \DeltaNu\ based only on a few modes around \nuMax\ gives rise to a  
locally defined $\epsilon$ (the  offset  in the asymptotic approximation of acoustic modes) which carries a signature of the evolutionary state.
Hence this method infers the evolutionary state of the deep interior  from the outer regions of the stars which are sensitive to the  conditions in the deep interior \citep{jcd2014}.
In a development of \citet{bedding2011}, \citet{stello2013} determine the period spacings from pairwise differences in mode frequencies in the expected range of the dipole modes.
Subsequently, \citet{mosser2012,mosser2014} have developed their method \citep{mosser2011mm} to allow for greater automation and to provide reliable estimates of the asymptotic period spacing for a large sample of stars. 
Finally, their latest developments are presented by \citet{mosser2015} and \citet{vrard2016} in which they apply a stretch to the power spectrum expressed in period to obtain a direct measure of the asymptotic period spacing.
Another automated method was presented by \citet{datta2015} based on the period \'echelle diagram to determine the asymptotic period spacings of red giants.
An alternative, comprehensive approach has been described by \citet{davies2016} who propose to make a complete analysis of the power spectrum including  the asymptotic period spacing and thereby determine the evolutionary state of the star.

These different methods are sensitive to different attributes of the spectrum. A comparison between the results from different methods will be provided in a subsequent paper. From this comparison it will be evident that all the methods find some stars difficult to classify and although the methods provide similar results for the majority of stars analysed, they provide different results for a non-negligible subset. 
In order to provide reliable results for a wide range of stars, it is therefore important to employ a range of different methods and to compare the determinations from the different methods.

In this paper we propose a new method to determine the evolutionary phase of HeB and RGB stars. This method is based on the width and structure of the mixed mode pattern and does not rely on an accurate measurement of the asymptotic period spacing itself. It is very efficient and requires minimal manual intervention making it suitable for the automated analysis  of large cohorts of stars.

\section{evidence for evolutionary state in the seismic spectrum}

\subsection{Key Indicators Used }
The problem being addressed is how, in  a highly efficient and autonomous way,  to identify the nuclear burning conditions in the centre of any given red-giant star using only the asteroseismic spectrum.

We start by considering  stars in different regions of the parameter space characterised by the location (\nuMax) of the peak of the power in the spectrum.
There are some values of the parameter \nuMax\ where it is unambiguously clear from theory that the star must be an RGB but there are other values where  no such  distinction is feasible.
We use BASTI models \citep{pietrinferni2004} with a margin for uncertainty to suggest that the only possible valid classification  for a red giant is RGB if \nuMax\ is below \LL\ or  above \UL. 
It should be remembered that no attempt is being made here to identify AGB stars.
In order to differentiate between red-clump and secondary-clump stars we also use \nuMax. Using theoretical work (BASTI tracks) and existing observations \citep[e.g.][]{mosser2014} to guide the choice, we have placed the boundary in \nuMax\ between the red clump and the secondary clump at \SC. 
We recognize that this is too firm a dividing point \ch{and that there is in reality no firm boundary as a function of \nuMax . It also} takes no account of the effect of metallicity and is something that we hope to improve in later versions of the algorithm. The mass of the star computed from asteroseismic parameters and spectroscopic temperature is used \textit{post hoc} to check that the initial evolutionary classification of the star is consistent with its mass.
We know that metallicity plays a large role. At low metallicity, even quite low-mass stars can become secondary-clump stars and, conversely,  at high metallicity,  stars of high enough mass to be in the secondary clump become red-clump stars. In this paper, the limit we chose is merely for guidance and this is an instance where comparison between different methods will be advantageous. 

As indicated earlier,  red-giant-branch stars and core-Helium-burning stars can be distinguished if the spacings in period of their mixed modes  can be determined.  However, it can be difficult to isolate individual modes in the spectrum and robust methods of fitting all the features including rotation are \ch{time-consuming} and still being developed.
We present a new method  that does not require individual modes to be isolated but instead uses all significant features in the spectrum.
The method relies on evaluating the differences in both the frequency and period domains between individual significant features of the $\ell=1$ modes. We show that there are characteristic patterns that can be used to identify the evolutionary state.

It is important that the  odd- and even-degree modes are considered separately and hence the first step in the process is to pick out the zones of the spectrum containing  the  odd- and even-degree modes in an automated manner.
To achieve this separation, we need the \ch{values} of \DeltaNu\ plus \eps\  which allow the zones to be delineated.
Here, we use the power spectrum of the power spectrum   \citep[e.g.][]{hekker2010}  and take the \DeltaNu\ value from  this so-called second spectrum.
We can accommodate errors  in the values of \DeltaNu\  as their effect is to produce small changes in the \eps\ value for which we can compensate. 
In order to  determine an optimal \eps, we  cross correlate a smoothed version of the spectrum over about six orders with the predictions of the \UP\ using the \DeltaNu\ obtained from second spectrum. The degree of smoothing used is  2\% of the value of \DeltaNu.
If the value of \DeltaNu\ is significantly wrong then there will be poor agreement with the \UP\ and the synchronism with the odd and even mode regions will be wrong. In this circumstance the \DeltaNu\ value must be improved. \ch{The code itself does not currently identify this need and in this case a mis-classification often occurs.}

With the values of \DeltaNu\ and \eps\, we can now divide the spectrum into two zones, one of which contains all the even$-\ell$~modes and the other all the odd$-\ell$~modes.
Because, in general, the initial estimate of \DeltaNu\  has good accuracy, this delineation of the zones works well in an autonomous manner.

\begin{figure}
\includegraphics[width=0.49\textwidth]{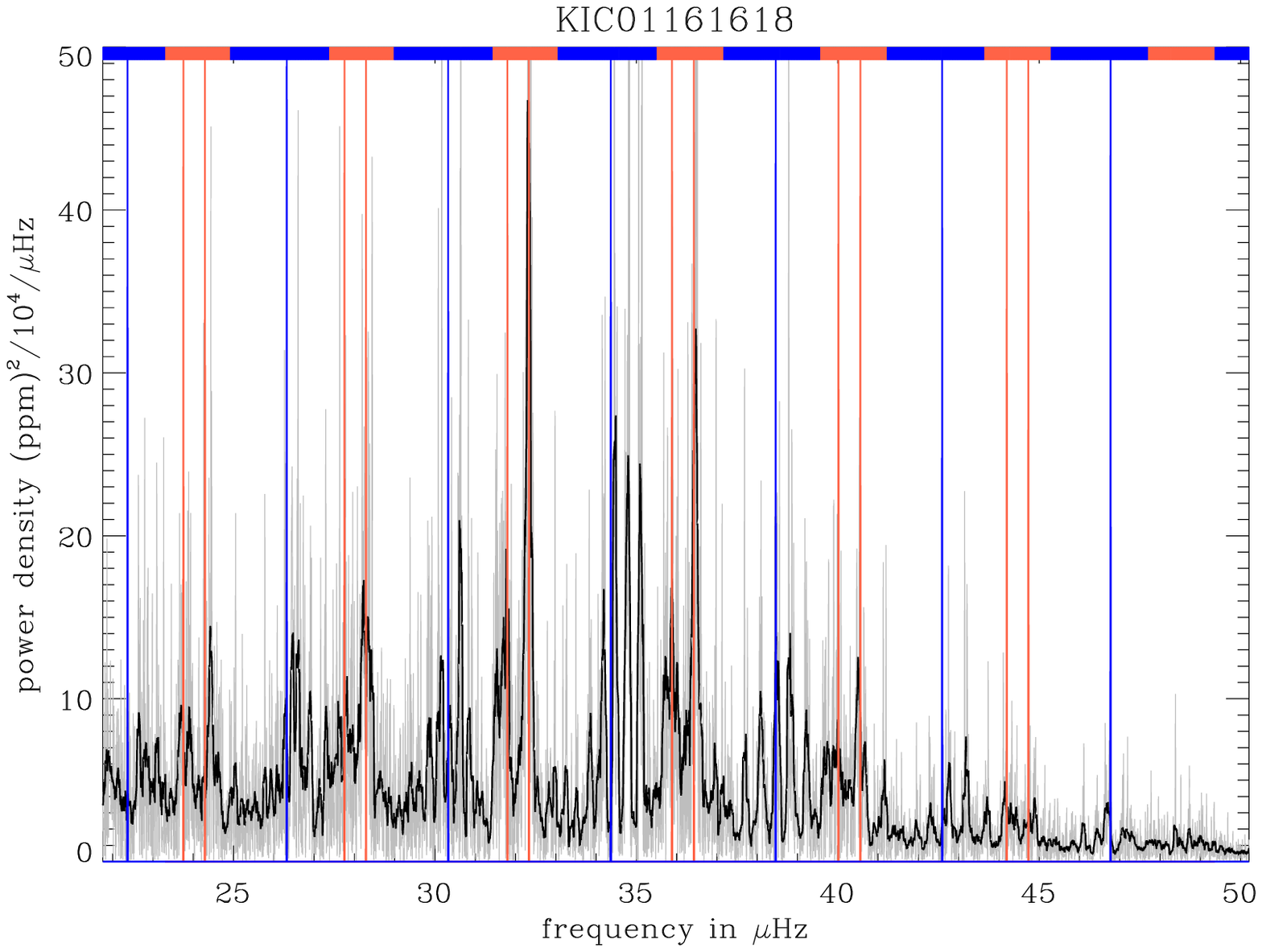}
\includegraphics[width=0.49\textwidth]{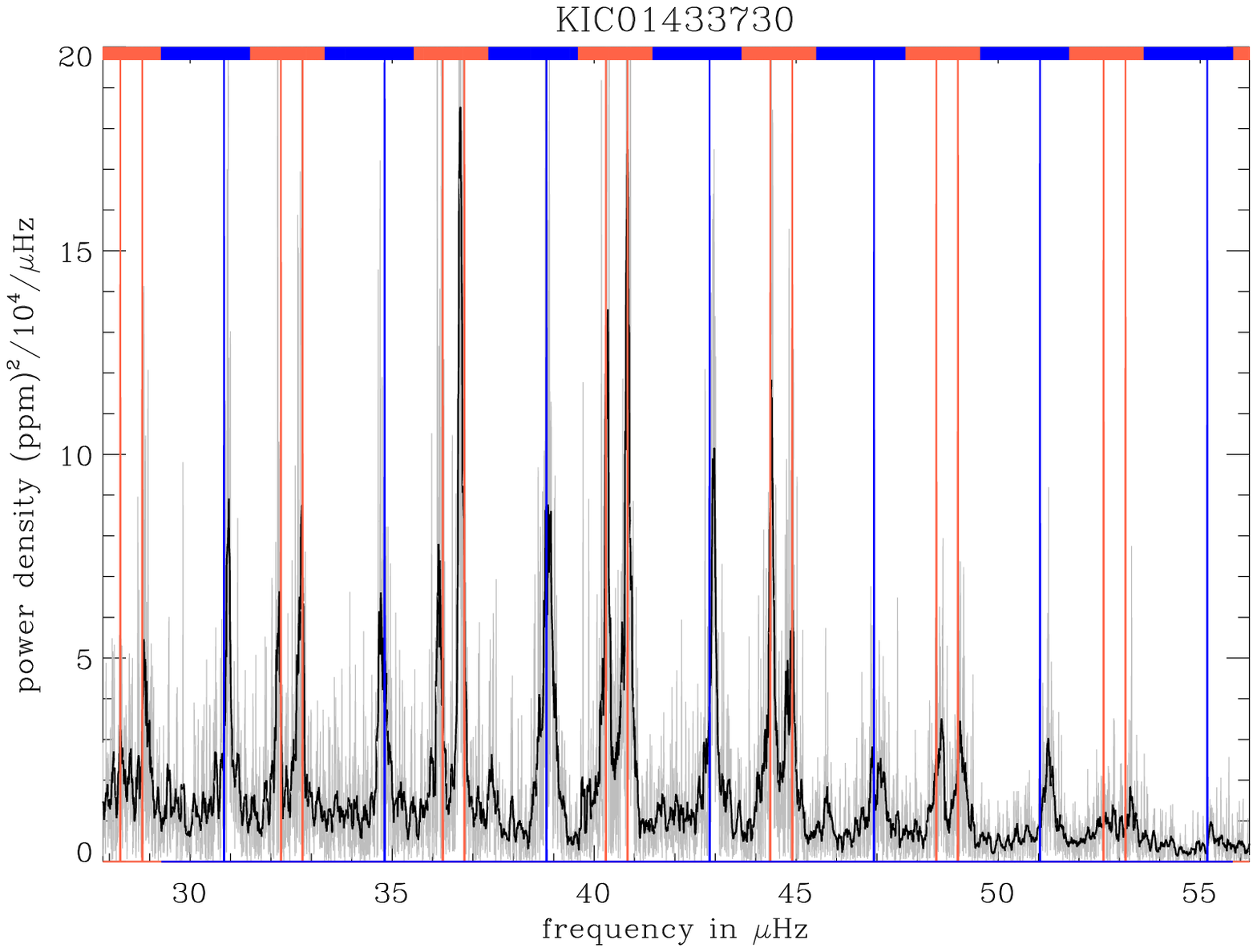}
\caption{The raw (in gray), the smoothed spectrum (in black) and the aligned Universal Pattern (in red/blue) for star KIC01161618 (red clump, top) and for star KIC01433730 (RGB, bottom). The coloured bar at the top of each plot gives an indication of where the (blue) odd- and (red) even$-\ell$ zones are. The vertical, coloured lines are predictions from the \UP.}
\label{UP}
\end{figure}

In Figure~\ref{UP}, we show, for two stars, the raw spectrum, the smoothed spectra, an indication of the locations of the odd and even zones  and the predictions from the \UP. The location of the \UP\ is  after the application of the small shift  determined from the cross correlation.
The narrow bar at the top of each plot indicates the locations of the zones.
All the  even$-\ell$~modes are in a zone  marked in red and all the odd$-\ell$~modes are in  zone  marked in blue.
The upper panel shows a HeB star and the lower one shows an RGB star. 
For more details on the dataset used in this paper see Section~\ref{data}.

%add in a figure which shows this stage

In each zone (i.e containing all the odd- or even- $\ell$~modes), we use statistical tests on the unsmoothed spectrum to pick out significant spikes \citep[e.g.][]{amb2007}. The significance threshold is set so as to exclude almost all the background noise but to pick up small features in the spectrum.
The background to the modes, against which a feature is tested, is determined according to the prescription 
given by \citet{kallinger2014} with an additional  multiplicative adjustment to bring the prediction into agreement with the  spectrum.
This is a frequentist approach that makes no assumptions about what will be present in the data. The result of the test is a set of frequencies at which spikes have been found.
The criterion employed is that a feature is considered significant if there is a less than 20\% probability of a false detection over a frequency range  of\uHz{3.5}. 
This range represents a typical \DeltaNu\ for a red-clump star.

In each zone (odd and even $\ell$), we find the separation in frequency of each feature from all other features. Thus, if there were $n$ features identified from the statistical test  there would be $n(n-1)$ frequency differences.  The absolute values of the frequency differences are used to form histograms -- one for each zone. The features to be found in these histograms carry information about the features within the spectrum and are different in the two zones of odd and even-$\ell$.

\begin{figure}
%these figures are produces by evol-YE-graphsA.pro using text files produced from the find evol code
\includegraphics[width=0.49\textwidth]{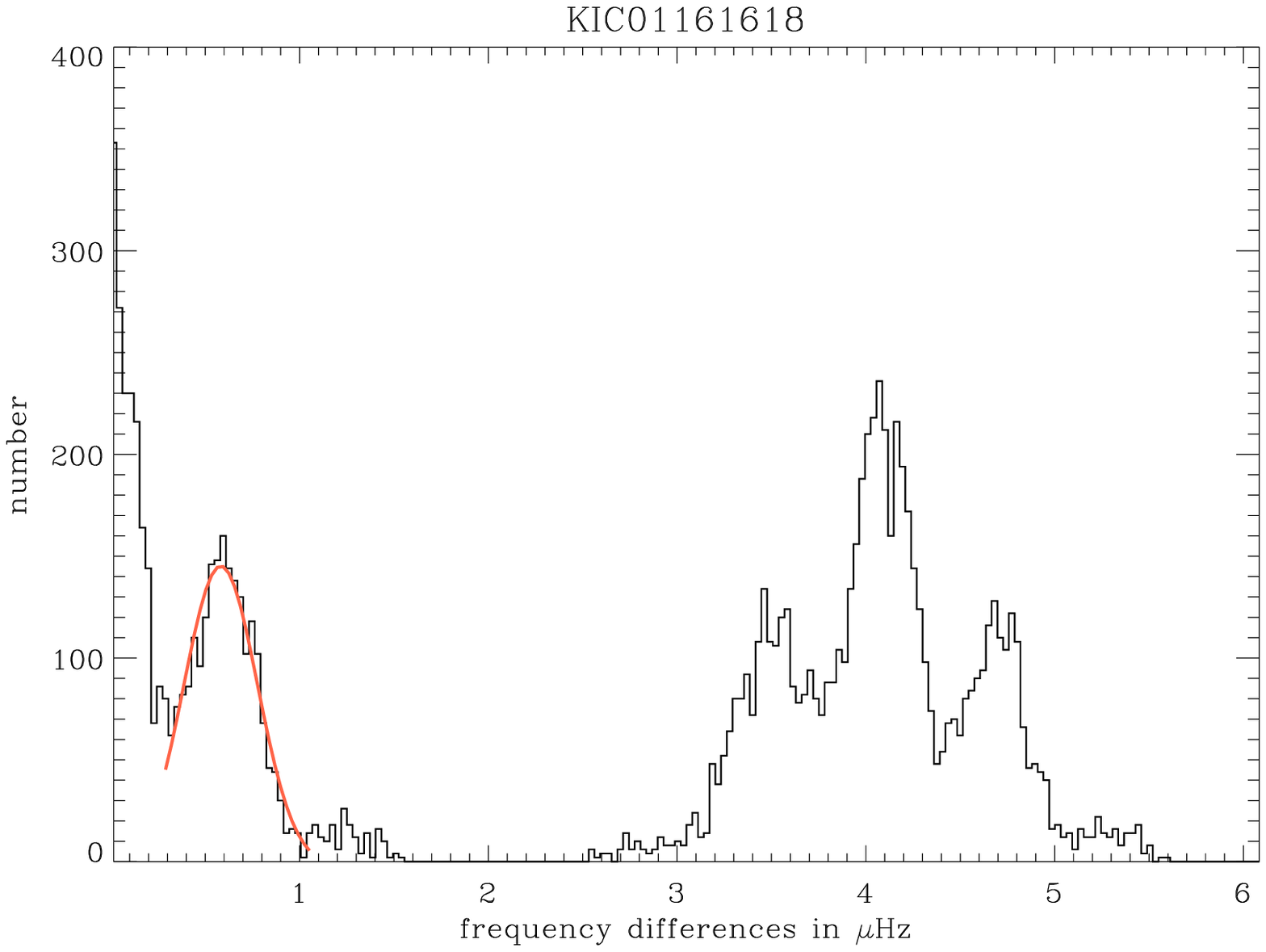}
\includegraphics[width=0.49\textwidth]{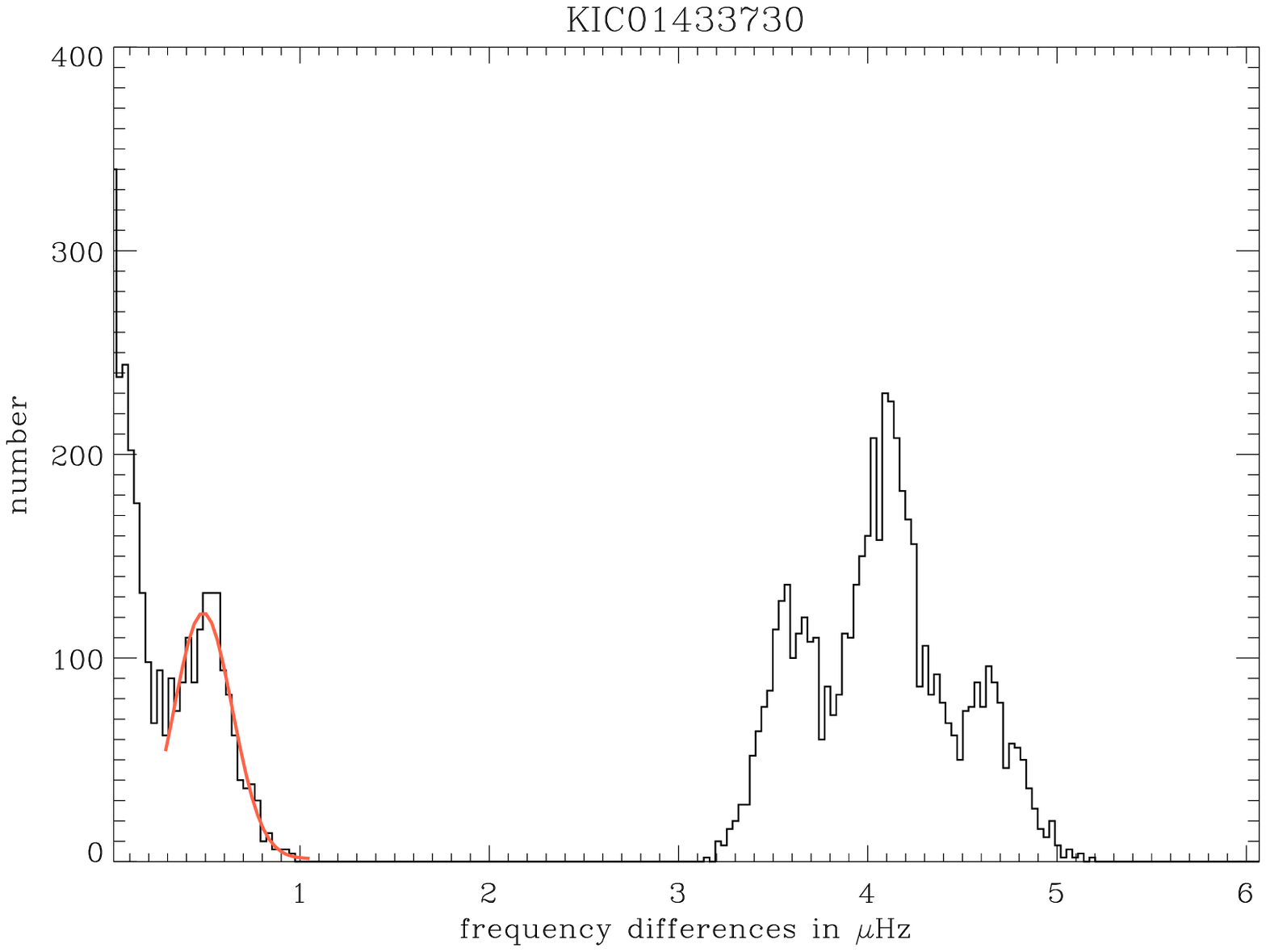}
\caption{Histograms of the absolute frequency differences for the even-$\ell$ zone for the same stars as in Figure~\ref{UP}. A Gaussian fit to
the $\delta\nu_{02}$ peak is shown in red. See the text for a description of the morphology of these histograms.}
\label{hist-even}
\end{figure}

Although the determination of  the evolutionary state rests on the odd-$\ell$ modes, we first consider the even-$\ell$ modes because the features in the even-$\ell$ frequency-difference histogram are straight forward  to identify. The characteristic frequency differences that we expect to see are the small spacing, $\delta\nu_{02}$ between $\ell=0$ and $\ell=2$ and also the large frequency spacing ($\equiv$ \DeltaNu). The small separation will appear three times, once close to zero frequency shift and around  the large frequency separation. Two examples are shown in Figure~\ref{hist-even}. The width of the features is due to the line widths of the underlying modes and the variation in \DeltaNu\ with $\nu$ across the spectrum. A fit to $\delta\nu_{02}$ is shown in  the figure for indicative purposes but is not used here.

% figure which shows this stage

\begin{figure}
\includegraphics[width=0.45\textwidth]{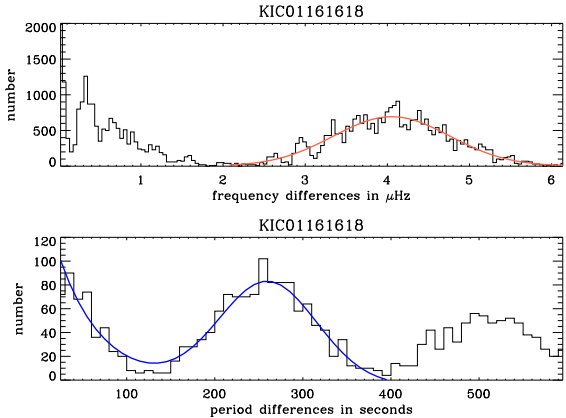}
\caption{Here we present  the data in the odd-$\ell$ zone only.
The upper panel  shows the histogram of the absolute frequency differences between any one feature and 
all others for the HeB star KIC01161618.  The Gaussian fitted to the data in the vicinity of \DeltaNu\ is shown in red. The lower panel  shows the same data but this time the histogram is composed from the absolute period differences. The blue curve is a multi-component fit to the data (see text for more details).}
\label{RC-histogram-all-f}
\end{figure}

\begin{figure}
\includegraphics[width=0.45\textwidth]{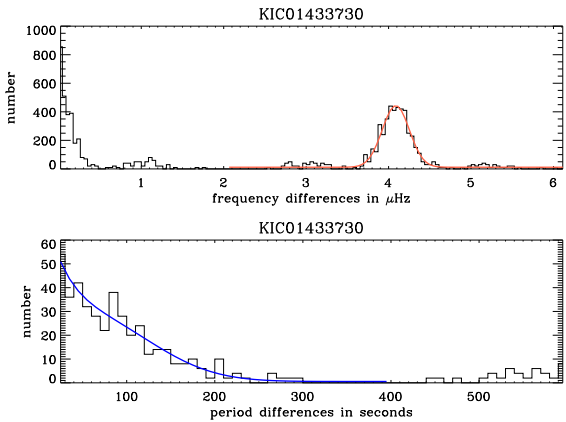}
\caption{Same as Fig.~\ref{RC-histogram-all-f} but now for the RGB star KIC01433730. }
\label{RGB-histogram-all-f}
\end{figure}

\begin{figure}
\includegraphics[width=0.45\textwidth]{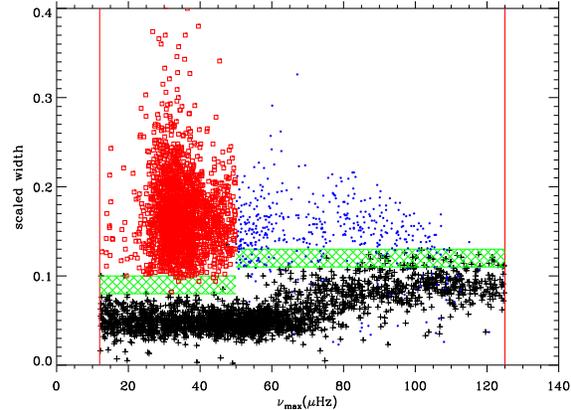}
\caption{Scaled width of the odd-mode frequency differences as a function of \nuMax\ showing the evolutionary categorisation of individual stars where black \ch{crosses are} RGB, red \ch{open squares} red clump and blue \ch{filled circles are} secondary clump. The ambiguous region is shown cross hatched in green.}
\label{graph-df-nmx}
\end{figure}

Histograms of the absolute frequency differences for the odd$-\ell$ zone  also show the large-frequency separation with additional features that are a consequence of the presence of mixed modes.
In the upper panels of Figure~\ref{RC-histogram-all-f} and Figure~\ref{RGB-histogram-all-f} we show the frequency-difference histograms for the same two stars as in Figure~\ref{UP} and Figure~\ref{hist-even}, this time for the odd-$\ell$ zone.
It is apparent that the width of the feature at \DeltaNu\ is very different for the two stars.
A Gaussian function is fitted to  the data around the initial value of \DeltaNu\ in order to quantify the width of the distribution and is shown as the red line in the figures.
\ch{The range over which the fit is taken is from 0.5*\DeltaNu\ to 1.5*\DeltaNu\ and is fully automatic.}
We divide the width of the Gaussian by the \DeltaNu\ value for the star to produce a scaled frequency width. For the whole cohort of stars considered here, we show this parameter as a function of \nuMax\ in Figure~\ref{graph-df-nmx}.

Before discussing the lower portions of  Figure~\ref{RC-histogram-all-f} and Figure~\ref{RGB-histogram-all-f}  and the detail of Figure~\ref{graph-df-nmx}  further, we  now consider how the mixed mode structure of an HeB star differs from that of an RGB star and how this is manifested in the difference histograms.

%add in a figure which shows this stage

HeB stars tend to show many mixed modes of similar heights over several p-mode orders with observed period spacings that typically lie between about 100\,s and 250\,s \citep[e.g.][]{bedding2011}. For RGB stars, the observed period spacing is much smaller at around 60\,s and the mixed modes have a sufficiently large inertia that few are visible around each nominal p mode \citep[e.g.][]{dupret2009}.
The two stars shown in Figure~\ref{UP} illustrate the extra structure present in HeB stars around the dipole modes. In Figure~\ref{UP}, the locations of the dipole modes are indicated by the blue lines from the Universal Pattern, for example, 
around\uHz{33} in the upper spectrum for KIC01161618.
This extra structure influences the features in the frequency-difference histogram.
Furthermore, by definition, the asymptotic period spacing is  uniform in period and not in frequency. This means that the frequency difference between two mixed modes at the low-frequency end of the spectrum is significantly smaller than the difference at the high-frequency end.

The widths 
of the  Gaussian fit to the features in the odd-$\ell$ zone, shown in red in the top panels of Figs~\ref{RC-histogram-all-f} and \ref{RGB-histogram-all-f}, expressed as a function of \DeltaNu\, can be used to differentiate the two classes of red giant.  To see why this is to be expected consider just two mixed modes at frequencies $\nu_1$ and $\nu_2$ separated by an interval corresponding to a given period spacing ($\Delta P$).
To a first approximation, the frequency separation between these modes is the product of the period spacing and the square of the mean frequency ($\nu_0=0.5(\nu_1+\nu_2$)) of the two features.

\begin{equation}
|\nu_1-\nu_2|=\Delta P \nu_0^2
\end{equation}

We can use this expression to give a measure of the variation in the observed frequency separation
for a set of modes at different frequencies, all of which have nearby features a fixed period interval away.
The  variation ($W_\nu$) in the frequency separation between the lowest ($\nu_{\rm low}$) and the highest ($\nu_{\rm high}$)  frequency  is given by

\begin{equation}
W_\nu=\Delta P(\nu_{\rm high}^2-\nu_{\rm low}^2)
\end{equation}

Assuming that the region of the spectrum considered is roughly symmetrical around the frequency of maximum oscillation power (\nuMax), and scaling by \DeltaNu\ to give a scaled width, we can re-write this expression as

\begin{equation}
\frac{W_\nu}{\Delta \nu}= \Delta P (\nu_{\rm high}-\nu_{\rm low}) \frac{2 \nu_{max}}{\Delta \nu}
\end{equation}

The width of the Gaussian fit is a measure of $W_\nu$.
Hence, for two stars at the same \nuMax\ but with different asymptotic period spacing, this simplified analysis predicts that the scaled width of the Gaussian fit to the  frequency differences for RGB stars should be several times smaller than the scaled width for an HeB star at the same \nuMax.

\begin{figure*}% the * makes it go over two columns
\includegraphics[width=0.9\textwidth]{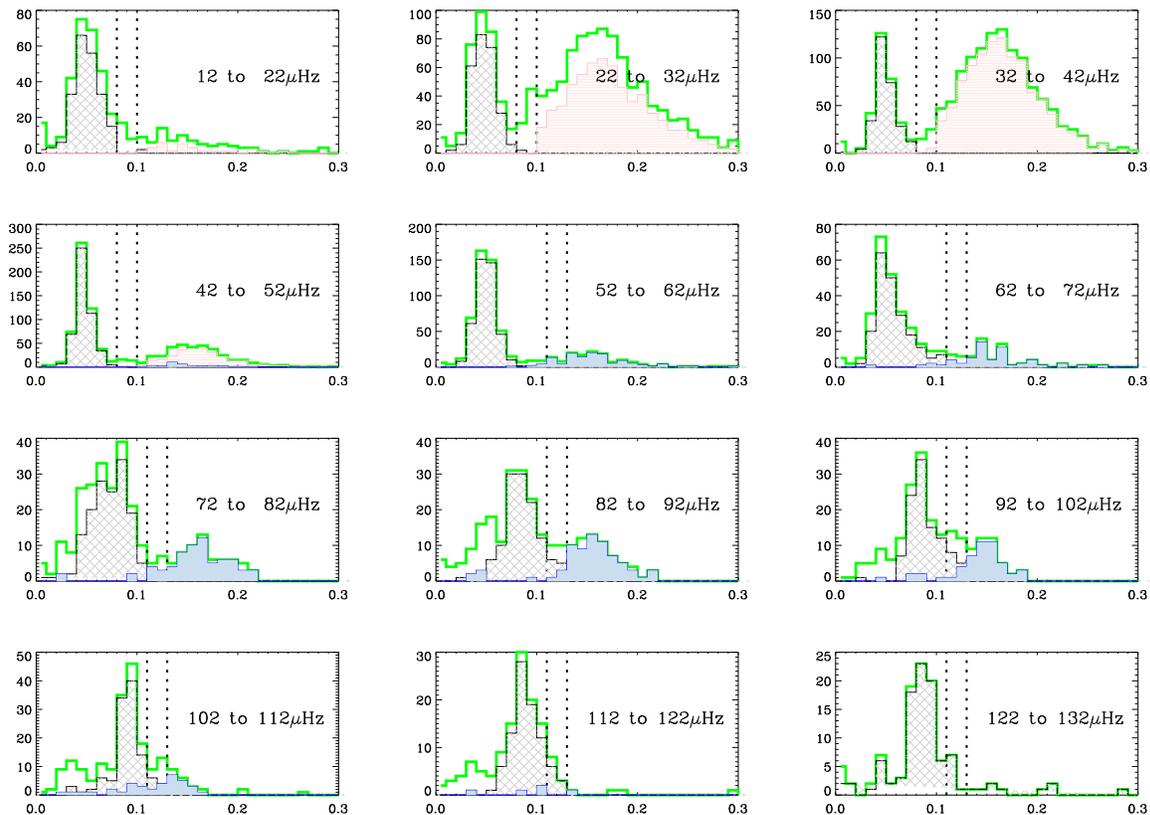}
\caption{Histograms of scaled width of the odd-mode frequency differences in \nuMax\ bands.
The vertical axis of each panel  is the number of stars and the horizontal axis is the scaled frequency width. The range of \nuMax\ considered is indicated on each panel. The colour-coding shows the determined evolutionary state as discussed in the text where grey\ch{(crossed hatched)} is RGB, red \ch{(horizontal lines)} is red clump and blue \ch{(solid)} is secondary clump. The green line gives the total number of stars in each bin including unclassified stars. The vertical, dashed, \ch{black} lines indicate the region where the scaled width provides no information of the evolutionary state.}
\label{freqwidth-nuMax}
\end{figure*}

The choice of the limits of the scaled width required to distinguish the evolutionary states is guided by the theory previously given, but is to some extent empirical. The limits are also dependent on the \nuMax\ value of the star.  To determine the values of the scaled width parameter corresponding to the different evolutionary states we show in Figure~\ref{freqwidth-nuMax} histograms of the values obtained for the scaled widths in different frequency bands.  It is evident that there are two distributions present in each frequency band.
Up to about\uHz{60}  there is a separation between the distributions such that above  a scaled frequency width of 0.1 the star is   a HeB star and below a relative width of 0.08 it is an RGB star.
In between these values there is some ambiguity which needs extra information to be resolved.

At higher frequencies the ambiguous region is at slightly higher scaled widths and there is less clear demarcation between the two distributions. 
For convenience, we call the frequency range (for \nuMax ) in which a star has the possibility to be a red-clump star  the  \IL\ region
and  the \IH\ region is where the star could be a secondary-clump star.

For the \nuMax\ range where a decision as to whether a star is in the RGB phase has to be made, Figure~\ref{graph-df-nmx} shows the scaled frequency widths for the individual stars   as a function of \nuMax.
The ambiguous region is shown cross hatched in green. 
As quantified in Table~\ref{numbers}, there are relatively few stars that fall into this ambiguous region.

Additionally, in the \IH\ region, it is possible to confuse RGB and secondary-clump stars and the classification is sometimes changed 
on the basis of secondary criteria.
These stars are evident in Figure~\ref{graph-df-nmx} as blue crosses, indicative of secondary-clump stars, in a region that is dominated by RGB stars.
We  use  as our supplementary parameters the value of the observed period separation of the mixed modes and the shape of the period-difference histogram.
We now describe the period-difference histogram.

A period-difference  histogram is formed in the same way as  the frequency-difference histogram. Again, the shapes of these histograms are sensitive to whether the star is an RGB or a HeB. The lower part of Figure~\ref{RC-histogram-all-f} shows the absolute 
period-difference histogram for the red-clump star KIC01161618. The ordinate in this plot is seconds and there is a broad feature peaked at about 270\,s that can be associated with the observed period spacing.
Stars which are identified as HeB stars from the signature in the frequency histogram often show period-difference histograms that are similar to this.

The lower part of Figure~\ref{RGB-histogram-all-f}  shows the period-difference histogram for the RGB star KIC01433730. The shape of this plot is different from that in the lower part of Figure~\ref{RC-histogram-all-f}. The general trend in the curve is a downward slope. There are some undulations which are due to the median period spacing but they do not dominate in the manner seen in  the HeB curve.
There are several contributions to this difference.
The nominal $\ell=1$ p mode is expected to be resolved and have a line width of around\uHz{0.1} (\citet{vrard2016}).
At a frequency of\uHz{30} this corresponds to a period width of about 100\,s and is responsible for the roughly linear fall off at low frequency in both the HeB and RGB plots.
Further broadening of the RGB plots comes from the mixed modes themselves. The expected RGB period spacings are small (perhaps 60\,s) and the inertia  curves are narrow  \citep[e.g.][]{dupret2009,grosjean2014} meaning that the amplitude of the modes falls off  steeply away from the position of the nominal p mode.  The situation is further complicated by the presence of rotation which gives splittings that can be comparable to the period spacings \citep{Mosser2012rot}. For HeB stars the rotational splitting is smaller and the period spacing larger leading to a clearer separation between the two phenomena.
 The nett consequence of all these for the RGB stars is that there is a wide range of observed spacings and the histogram does not show a clear peak but instead a broad triangular distribution with modulation due to the period spacing.

The different morphology between the two types of period-difference histograms is very apparent visually and we fit an empirical function to the period differences in order to automate the detection of the distinction. The function we choose is composed of  decaying exponentials and an offset  Gaussian plus a constant term. We test the coefficients  to see if the addition of the Gaussian and/or more than one exponential decay is warranted using the Bayesian Information Criterion \citep[BIC, ][]{schwarz1978}.
For the two stars in Figure~\ref{RC-histogram-all-f} and Figure~\ref{RGB-histogram-all-f}, the result of the fits is shown by the blue line.
Hence, from the need for a Gaussian term and the location of its peak it is possible to distinguish between the two different shapes of the period-difference histograms. The peak of the Gaussian, if significant, is used as an estimate of the observed mixed mode period spacing. In the discussion that follows, we refer to this as the mode (in the statistical sense of most likely) observed period spacing or, more briefly, as mode-\OPS .

\subsection{The Ambiguous Region and Identifying the secondary clump}
\label{AZ}
A parallel statistical analysis is conducted which isolates individual modes, estimates their central frequency from the location of the peak power, and directly provides a number of values for the  observed period spacing for the $\ell=1$ modes. The median of these values is used as an estimate of the observed period spacing. 
In the discussion that follows, we refer to this as the median observed period spacing or, more briefly, as median-\OPS . 

For stars in the ambiguous region where the scaled frequency widths do not provide a decision and also to distinguish secondary clump from RGB, we use values of the estimates of the observed period spacing.
We have the two measures of the observed period spacing; one (median-\OPS ) comes from the separation of individual modes and the other (mode-\OPS ) from the fit to the period histogram of the prominent spikes. If both values are below \PL\ then the star is placed in the RGB category.
If the two measures of the observed period spacing are above \PL\ then we consider that the star cannot be on the RGB and is
secondary clump or red clump.
If the two measures of observed period spacing do not agree, then the star is unclassified.

As can be seen from Figures~\ref{graph-df-nmx} and~\ref{freqwidth-nuMax}, the scaled frequency width is not as clear  a distinguishing parameter for the secondary clump vs. RGB as it is for red clump vs. RGB. We therefore test the observed period spacing for all the RGB stars with \nuMax\ above \SC\ to check that they have the expected, low period spacing. This check leads to a small number of the stars being re-classified from RGB to secondary clump. 
If the two measures of observed period spacing do not agree, then the star is unclassified.
However, we find that an original classification as secondary clump (i.e. an above threshold value for the scaled frequency width) is robust and in this case no further checks are made.

\subsection{Stars with low $\ell=1$}
The method described here depends on the presence of enough signal in the $\ell=1$ modes for the feature to be analysed.
There are situations  where the $\ell=1$ modes are of low  amplitude or altogether missing 
\citep{mosser2012,garcia2014,stello2016}.  Although we do not measure the mode linewidth here and hence cannot directly convert the mode height to  mode amplitude, we do need to be aware of the phenomenon.
Under the assumption that the mode linewidth does not vary strongly from star to star, this situation can be picked up automatically by checking the height of the detected modes with respect to the local background. Typically, we use a signal-to-noise ratio of less than about 15 to indicate modes of low $\ell=1$ height.

\section{Data}
\label{data}
The data used in this paper were taken by instrumentation on the \textit{Kepler} satellite \citep{borucki2010}. 
We give no detail of the data collection here except to say that we have used solely the long cadence data taken with a sampling time of about 30\,minutes. The data were prepared according to the principles discussed in \citet{handberg2014}. 
The subset of stars analysed here form part of the set of stars analysed by APOKASC collaboration whose purpose is to consider stars whose asteroseismic data is provided from \textit{Kepler} and with spectroscopic data from the high-resolution Apache Point Observatory Galactic Evolution
Experiment (APOGEE)\citep{gunn2006} spectra analyzed by members of the
third Sloan Digital Sky Survey (SDSS-III). A subset of these stars have been  analysed previously and the results are reported in \citet{pinsonneault2014} where more details  of the project and the methods used to select the stars can be found.

The spectroscopic data for temperature and metallicity used in some of the plots in this paper are provided by APOGEE data release 13. Information about the release and the data products are to be found at http://www.sdss.org/dr13/.

\section{Results}
\label{results}
The stars considered here are some 6637 stars from the APOKASC set.  Of these, 33 have \nuMax\ values  above the frequency at which  it  is considered to be too close to the Nyquist cut off for a reliable determination of \nuMax. \ch{For a discussion of some of the issues associated with stars whose \nuMax\ is close to the Nyquist cut off see \citet{yu2016}.} Given the high \nuMax, these stars must be RGB stars but the nature of their spectra means that we cannot process them with this method. This leaves 6604 for which we aim  to provide  a categorisation. We will consider the stars in four bands which are \textit{low} = below the lower limit of \LL\ for red-clump stars, \textit{high} = above the limit of \UL\ for secondary-clump stars, and two intermediate bands called \IL\ and \IH\ where the red-clump and secondary-clump stars will be placed. In Panel E of Table~\ref{numbers}  we give the final numbers in each category for the stars. The remaining panels are indicative of some of the stages in the process.

\begin{table}
\centering
\caption{The numbers of stars in the various categories and evolutionary states.
The notation used for the various evolutionary states is  RC=red clump, SC=secondary clump, U=unclassified.}
Panel A: Numbers of stars in the different frequency bands.\\ [1ex]
\label{numbers}
\begin{tabular}{|l|c|c|c|c|}
\cline{2-5}
\multicolumn{1}{c|}{}   &\textit{low}	&\IL	&\IH	&\textit{high}\\
\hline
numbers of stars	&475					&3482	&1982	&665	\\
\% of stars				&7.2					&52.7	&30.0	&10.1\\ [1ex]
\hline
\end{tabular}
\bigskip

Panel B: Percentage of stars  in \IL\ and \IH\ bands assigned directly by the scaled frequency width, in the ambiguous region  and  unclassified because of low $\ell=1$.\\ [1ex]
\begin{tabular}{|l|c|c|c|}
\cline{2-4}
\multicolumn{1}{c|}{}   	&\IL	&\IH	&\textit{high}\\
\hline
direct				&90.3		&85.7	&-	\\
ambiguous			&2.1		&3.7	&-	\\
low $\ell=1$	&7.6		&10.6	&17.0\\ [1ex]
\hline
\end{tabular}
\bigskip

Panel C: How the ambiguous region gets classified. The columns give the numbers of stars in each category. \\ [1ex]
\begin{tabular}{|l|c|c|}
\cline{2-3}
\multicolumn{1}{c|}{}   &\IL	&\IH	\\
\hline
RGB				&1	&12		\\
RC/SC			&39	&44	\\
U	&35		&17	\\ [1ex]
\hline
\end{tabular}
\bigskip

Panel D: Number of stars in the \IH\ range whose classification is changed because of the observed period spacing.\\ [1ex]
\begin{tabular}{|l|l|}
\hline
RGB $\rightarrow$ SC										&	58\\
classified as RGB $\rightarrow$ unclassified & 17\\ [1ex]
\hline
\end{tabular}
\bigskip

Panel E: Final classification -- the numbers of stars in each frequency band and evolutionary state.\\ [1ex]
\begin{tabular}{|l|c|c|c|c|}
\cline{2-5}
\multicolumn{1}{c|}{}   &\textit{low}	&\IL	&\IH	&\textit{high}\\
\hline
RGB		&475&	1107&	1319&	665\\
RC/SC	&-	&	2077&	439&	-\\
U			&-	&	298&	224&	-\\ [1ex]
\hline
\end{tabular}
\end{table}

\begin{itemize}
\item Panel A gives the numbers and percentages of the cohort  of 6604 stars in the four frequency bands.
\item Panel B gives the percentages of stars that are directly assigned a classification, are in the ambiguous region and that have too low an amplitude in the $\ell=1$ to be classified using the $\ell=1$ modes. The percentages are computed for each band as given in Panel A.
\item Panel C gives the classification of the stars in the ambiguous region.
\item Panel D gives the reclassification in the \IH\ range because of the observed period spacing.
\item Panel E give the final classification numbers for the stars.
\end{itemize}

The individual classifications are provided in Table~\ref{tableOfResults} which gives the KIC number and the classification for each of the 6637 stars considered here including those with a \nuMax\ too close to the Nyquist frequency to be processed. 

%the very big table
\begin{table}
\centering
\caption{Classification for all the stars analysed. In column 1 is the KIC number of the star concerned. 
In column 2 is the classification where RGB=red-giant branch, RC=red clump, 2CL=secondary clump,U=unclassified and nmx-too-high=\nuMax\ is above the range considered. Note that only the first 20 lines provided here and the rest are available on line.}
\label{tableOfResults}
\begin{tabular}{|l|l|}
\hline
KIC number & Classification\\
\hline
\input{shortTable.txt}
\hline
\end{tabular}
\end{table}

\begin{figure*}
\includegraphics[width=0.75\textwidth]{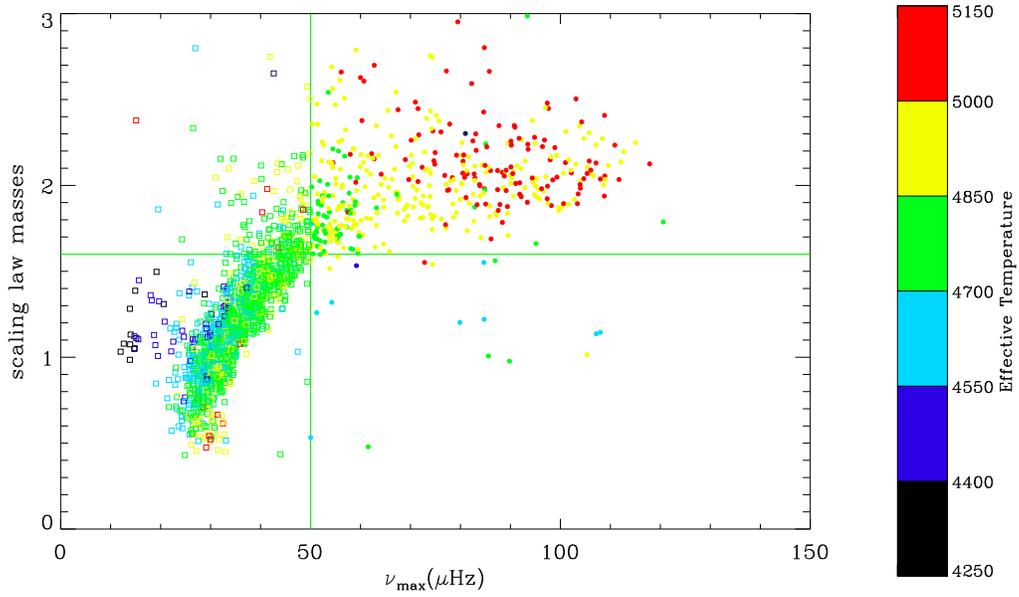}
\caption{The scaling-law mass in units of a solar mass as a function of \nuMax\ for the stars categorised as red clump (shown in open squares) and secondary clump (shown in filled-in circles). The data points are colour coded by effective temperature in kelvin provided by APOGEE DR13 (see Section~\ref{data}). The horizontal green line is drawn at 1.6 solar masses. The vertical line at \SC\ is the nominal division point between red clump and secondary clump.}
\label{HeB-T}
\end{figure*}

\begin{figure*}
\includegraphics[width=0.75\textwidth]{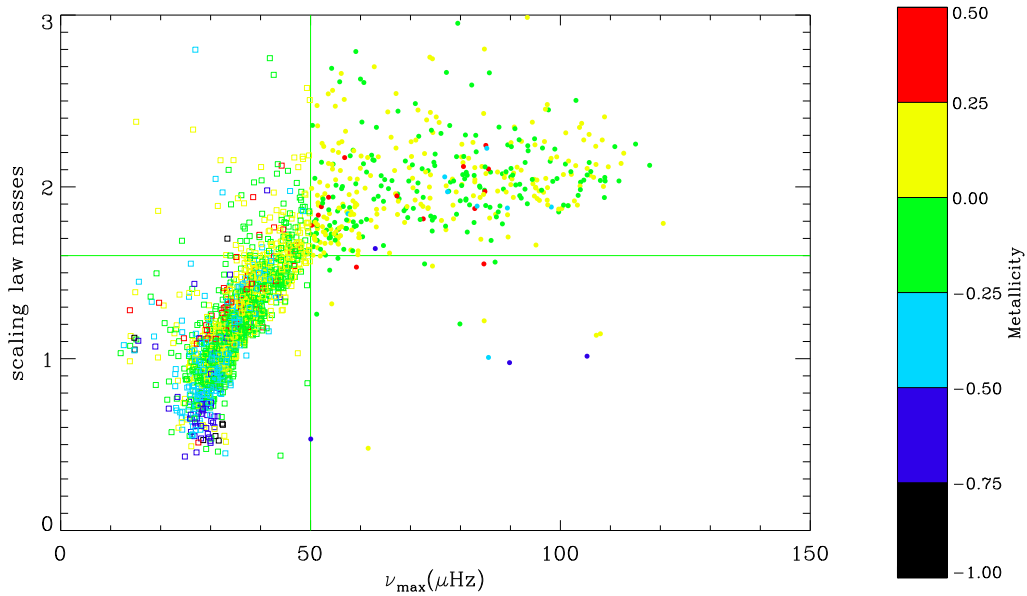}
\caption{Same as Fig.~\ref{HeB-T}, but now with the data points colour coded by metallicity.}
\label{HeB-Z}
\end{figure*}

For the red-clump and secondary-clump stars, it is informative to look at how the masses derived from the scaling laws \citep{kb1995} fit with the classifications. These are shown in Figures~\ref{HeB-T} and~\ref{HeB-Z} which give the scaling-law mass as a function of \nuMax\ colour coded for effective temperature and metallicity respectively. As expected, the low-metallicity, red-clump stars are to be found at lower than average masses and higher than average temperatures. Most of the secondary-clump stars are above 1.7 solar masses and they are among the hotter stars. There are some high mass, secondary-clump stars which need  further investigation which is outside the scope of this paper. They may be genuinely RGB stars or they may have incorrectly determined  asteroseismic parameters.

\section{Discussion and Conclusions}
\label{discussion}
A robust method to determine the evolutionary state of red-giant stars has been presented here.
This will be an important factor in the determination of the ages of stars in stellar populations.
The method uses the morphology of the power spectrum to do the classification. 
As long as sufficient power is present in the dipole modes, the method is fast and reliable.  Additionally, the method provides an estimate of the observed period spacing. 
Although there is some uncertainty in the boundary between the red clump and the secondary clump, for many purposes it is sufficient to know that the star is in its core-Helium-burning phase. Furthermore, the available spectroscopic information together with stellar models may be used to resolve the ambiguity.

The method has been used within the APOKASC consortium  as part of the process to classify the stars. 
A paper that details the processes used to  combine several different classification methods to provide the overall classification is in preparation.


\section*{Acknowledgments}
YE and GRD acknowledge the support of the UK Science and Technology Facilities Council (STFC).
SH has received funding from the European Research Council under the European Community’s Seventh Framework Programme
(FP7/2007-2013) / ERC grant agreement no 338251 (StellarAges).
SB acknowledges partial support of NASA grant NNX16AI09G and NNX13AE70G and NSF grant AST-1514676.
Funding for the Stellar Astrophysics Centre (SAC) is provided by The Danish National Research Foundation (Grant agreement no.: DNRF106).
%\bibliographystyle{mn2e_new}
%\bibliographystyle{plain}
%\bibliography{evol-v7}

\label{lastpage}

\end{document}

%% file: shortTable.txt
1027337 &   RGB   \\
1160789 &   RC   \\
1160986 &   RGB   \\
        1161447 &   RC   \\
        1161618 &   RC   \\
        1162220 &   RGB   \\
        1162746 &   RC   \\
        1163114 &   RGB   \\
        1163359 &   RGB   \\
        1163621 &   2CL   \\
        1294122 &   RGB   \\
        1294385 &   RGB   \\
        1296068 &   RGB   \\
        1296507 &   U   \\
        1429505 &   RGB   \\
        1431059 &   RGB   \\
        1431599 &   RGB   \\
        1432587 &   RGB   \\
        1433593 &   U   \\
        1433730 &   RGB  \\

%% file: evol-v7.bbl
\begin{thebibliography}{41}
\expandafter\ifx\csname natexlab\endcsname\relax\def\natexlab#1{#1}\fi

\bibitem[{{Beck} {et~al}\mbox{.}(2011){Beck}, {Bedding}, {Mosser}, {Stello},
  {Garcia}, {Kallinger}, {Hekker}, {Elsworth}, {Frandsen}, {Carrier}, {De
  Ridder}, {Aerts}, {White}, {Huber}, {Dupret}, {Montalb{\'a}n}, {Miglio},
  {Noels}, {Chaplin}, {Kjeldsen}, {Christensen-Dalsgaard}, {Gilliland},
  {Brown}, {Kawaler}, {Mathur}, \& {Jenkins}}]{beck2011}
{Beck} P.~G. {et~al.}, 2011, Science, 332, 205

\bibitem[{{Bedding} {et~al}\mbox{.}(2010){Bedding}, {Huber}, {Stello},
  {Elsworth}, {Hekker}, {Kallinger}, {Mathur}, {Mosser}, {Preston}, {Ballot},
  {Barban}, {Broomhall}, {Buzasi}, {Chaplin}, {Garc{\'{\i}}a}, {Gruberbauer},
  {Hale}, {De Ridder}, {Frandsen}, {Borucki}, {Brown}, {Christensen-Dalsgaard},
  {Gilliland}, {Jenkins}, {Kjeldsen}, {Koch}, {Belkacem}, {Bildsten}, {Bruntt},
  {Campante}, {Deheuvels}, {Derekas}, {Dupret}, {Goupil}, {Hatzes}, {Houdek},
  {Ireland}, {Jiang}, {Karoff}, {Kiss}, {Lebreton}, {Miglio}, {Montalb{\'a}n},
  {Noels}, {Roxburgh}, {Sangaralingam}, {Stevens}, {Suran}, {Tarrant}, \&
  {Weiss}}]{bedding2010}
{Bedding} T.~R. {et~al.}, 2010, \apjl, 713, L176

\bibitem[{{Bedding} {et~al}\mbox{.}(2011){Bedding}, {Mosser}, {Huber},
  {Montalb{\'a}n}, {Beck}, {Christensen-Dalsgaard}, {Elsworth},
  {Garc{\'{\i}}a}, {Miglio}, {Stello}, {White}, {De Ridder}, {Hekker}, {Aerts},
  {Barban}, {Belkacem}, {Broomhall}, {Brown}, {Buzasi}, {Carrier}, {Chaplin},
  {di Mauro}, {Dupret}, {Frandsen}, {Gilliland}, {Goupil}, {Jenkins},
  {Kallinger}, {Kawaler}, {Kjeldsen}, {Mathur}, {Noels}, {Aguirre}, \&
  {Ventura}}]{bedding2011}
---, 2011, Nature, 471, 608

\bibitem[{{Borucki} {et~al}\mbox{.}(2010){Borucki}, {Koch}, {Basri}, {Batalha},
  {Brown}, {Caldwell}, {Caldwell}, {Christensen-Dalsgaard}, {Cochran},
  {DeVore}, {Dunham}, {Dupree}, {Gautier}, {Geary}, {Gilliland}, {Gould},
  {Howell}, {Jenkins}, {Kondo}, {Latham}, {Marcy}, {Meibom}, {Kjeldsen},
  {Lissauer}, {Monet}, {Morrison}, {Sasselov}, {Tarter}, {Boss}, {Brownlee},
  {Owen}, {Buzasi}, {Charbonneau}, {Doyle}, {Fortney}, {Ford}, {Holman},
  {Seager}, {Steffen}, {Welsh}, {Rowe}, {Anderson}, {Buchhave}, {Ciardi},
  {Walkowicz}, {Sherry}, {Horch}, {Isaacson}, {Everett}, {Fischer}, {Torres},
  {Johnson}, {Endl}, {MacQueen}, {Bryson}, {Dotson}, {Haas}, {Kolodziejczak},
  {Van Cleve}, {Chandrasekaran}, {Twicken}, {Quintana}, {Clarke}, {Allen},
  {Li}, {Wu}, {Tenenbaum}, {Verner}, {Bruhweiler}, {Barnes}, \&
  {Prsa}}]{borucki2010}
{Borucki} W.~J. {et~al.}, 2010, Science, 327, 977

\bibitem[{{Broomhall} {et~al}\mbox{.}(2007){Broomhall}, {Chaplin}, {Elsworth},
  \& {Appourchaux}}]{amb2007}
{Broomhall} A.~M., {Chaplin} W.~J., {Elsworth} Y., {Appourchaux} T., 2007,
  \mnras, 379, 2

\bibitem[{{Christensen-Dalsgaard}(2014)}]{jcd2014book}
{Christensen-Dalsgaard} J., 2014, in Asteroseismology, {Pall{\'e}} P.~L.,
  {Esteban} C., eds., Cambridge University Press, p. 194

\bibitem[{{Christensen-Dalsgaard} {et~al}\mbox{.}(2014){Christensen-Dalsgaard},
  {Silva Aguirre}, {Elsworth}, \& {Hekker}}]{jcd2014}
{Christensen-Dalsgaard} J., {Silva Aguirre} V., {Elsworth} Y., {Hekker} S.,
  2014, \mnras, 445, 3685

\bibitem[{{Corsaro} {et~al}\mbox{.}(2012){Corsaro}, {Stello}, {Huber},
  {Bedding}, {Bonanno}, {Brogaard}, {Kallinger}, {Benomar}, {White}, {Mosser},
  {Basu}, {Chaplin}, {Christensen-Dalsgaard}, {Elsworth}, {Garc{\'{\i}}a},
  {Hekker}, {Kjeldsen}, {Mathur}, {Meibom}, {Hall}, {Ibrahim}, \&
  {Klaus}}]{corsaro2012}
{Corsaro} E. {et~al.}, 2012, \apj, 757, 190

\bibitem[{{Datta} {et~al}\mbox{.}(2015){Datta}, {Mazumdar}, {Gupta}, \&
  {Hekker}}]{datta2015}
{Datta} A., {Mazumdar} A., {Gupta} U., {Hekker} S., 2015, \mnras, 447, 1935

\bibitem[{{Davies} \& {Miglio}(2016)}]{davies2016}
{Davies} G.~R., {Miglio} A., 2016, ArXiv e-prints

\bibitem[{{De Ridder} {et~al}\mbox{.}(2009){De Ridder}, {Barban}, {Baudin},
  {Carrier}, {Hatzes}, {Hekker}, {Kallinger}, {Weiss}, {Baglin}, {Auvergne},
  {Samadi}, {Barge}, \& {Deleuil}}]{deridder2009}
{De Ridder} J. {et~al.}, 2009, \nat, 459, 398

\bibitem[{{Deheuvels} {et~al}\mbox{.}(2015){Deheuvels}, {Ballot}, {Beck},
  {Mosser}, {{\O}stensen}, {Garc{\'{\i}}a}, \& {Goupil}}]{deheuvels2015}
{Deheuvels} S., {Ballot} J., {Beck} P.~G., {Mosser} B., {{\O}stensen} R.,
  {Garc{\'{\i}}a} R.~A., {Goupil} M.~J., 2015, \aap, 580, A96

\bibitem[{{Dupret} {et~al}\mbox{.}(2009){Dupret}, {Belkacem}, {Samadi},
  {Montalban}, {Moreira}, {Miglio}, {Godart}, {Ventura}, {Ludwig},
  {Grigahc{\`e}ne}, {Goupil}, {Noels}, \& {Caffau}}]{dupret2009}
{Dupret} M.-A. {et~al.}, 2009, \aap, 506, 57

\bibitem[{{Garc{\'{\i}}a} {et~al}\mbox{.}(2014){Garc{\'{\i}}a}, {P{\'e}rez
  Hern{\'a}ndez}, {Benomar}, {Silva Aguirre}, {Ballot}, {Davies}, {Do{\u g}an},
  {Stello}, {Christensen-Dalsgaard}, {Houdek}, {Ligni{\`e}res}, {Mathur},
  {Takata}, {Ceillier}, {Chaplin}, {Mathis}, {Mosser}, {Ouazzani},
  {Pinsonneault}, {Reese}, {R{\'e}gulo}, {Salabert}, {Thompson}, {van Saders},
  {Neiner}, \& {De Ridder}}]{garcia2014}
{Garc{\'{\i}}a} R.~A. {et~al.}, 2014, \aap, 563, A84

\bibitem[{{Grosjean} {et~al}\mbox{.}(2014){Grosjean}, {Dupret}, {Belkacem},
  {Montalban}, {Samadi}, \& {Mosser}}]{grosjean2014}
{Grosjean} M., {Dupret} M.-A., {Belkacem} K., {Montalban} J., {Samadi} R.,
  {Mosser} B., 2014, \aap, 572, A11

\bibitem[{{Gunn} {et~al}\mbox{.}(2006){Gunn}, {Siegmund}, {Mannery}, {Owen},
  {Hull}, {Leger}, {Carey}, {Knapp}, {York}, {Boroski}, {Kent}, {Lupton},
  {Rockosi}, {Evans}, {Waddell}, {Anderson}, {Annis}, {Barentine}, {Bartoszek},
  {Bastian}, {Bracker}, {Brewington}, {Briegel}, {Brinkmann}, {Brown}, {Carr},
  {Czarapata}, {Drennan}, {Dombeck}, {Federwitz}, {Gillespie}, {Gonzales},
  {Hansen}, {Harvanek}, {Hayes}, {Jordan}, {Kinney}, {Klaene}, {Kleinman},
  {Kron}, {Kresinski}, {Lee}, {Limmongkol}, {Lindenmeyer}, {Long}, {Loomis},
  {McGehee}, {Mantsch}, {Neilsen}, {Neswold}, {Newman}, {Nitta}, {Peoples},
  {Pier}, {Prieto}, {Prosapio}, {Rivetta}, {Schneider}, {Snedden}, \&
  {Wang}}]{gunn2006}
{Gunn} J.~E. {et~al.}, 2006, \aj, 131, 2332

\bibitem[{{Handberg} \& {Lund}(2014)}]{handberg2014}
{Handberg} R., {Lund} M.~N., 2014, \mnras, 445, 2698

\bibitem[{{Hekker}(2013)}]{hekker2013}
{Hekker} S., 2013, Advances in Space Research, 52, 1581

\bibitem[{{Hekker} {et~al}\mbox{.}(2011){Hekker}, {Basu}, {Stello},
  {Kallinger}, {Grundahl}, {Mathur}, {Garc{\'{\i}}a}, {Mosser}, {Huber},
  {Bedding}, {Szab{\'o}}, {De Ridder}, {Chaplin}, {Elsworth}, {Hale},
  {Christensen-Dalsgaard}, {Gilliland}, {Still}, {McCauliff}, \&
  {Quintana}}]{hekker2011}
{Hekker} S. {et~al.}, 2011, \aap, 530, A100

\bibitem[{{Hekker} {et~al}\mbox{.}(2010){Hekker}, {Broomhall}, {Chaplin},
  {Elsworth}, {Fletcher}, {New}, {Arentoft}, {Quirion}, \&
  {Kjeldsen}}]{hekker2010}
---, 2010, \mnras, 402, 2049

\bibitem[{{Hekker} {et~al}\mbox{.}(2009){Hekker}, {Kallinger}, {Baudin}, {De
  Ridder}, {Barban}, {Carrier}, {Hatzes}, {Weiss}, \& {Baglin}}]{hekker2009}
---, 2009, \aap, 506, 465

\bibitem[{{Huber} {et~al}\mbox{.}(2010){Huber}, {Bedding}, {Stello}, {Mosser},
  {Mathur}, {Kallinger}, {Hekker}, {Elsworth}, {Buzasi}, {De Ridder},
  {Gilliland}, {Kjeldsen}, {Chaplin}, {Garc{\'{\i}}a}, {Hale}, {Preston},
  {White}, {Borucki}, {Christensen-Dalsgaard}, {Clarke}, {Jenkins}, \&
  {Koch}}]{huber2010}
{Huber} D. {et~al.}, 2010, \apj, 723, 1607

\bibitem[{{Kallinger} {et~al}\mbox{.}(2014){Kallinger}, {De Ridder}, {Hekker},
  {Mathur}, {Mosser}, {Gruberbauer}, {Garc{\'{\i}}a}, {Karoff}, \&
  {Ballot}}]{kallinger2014}
{Kallinger} T. {et~al.}, 2014, \aap, 570, A41

\bibitem[{{Kallinger} {et~al}\mbox{.}(2012){Kallinger}, {Hekker}, {Mosser}, {De
  Ridder}, {Bedding}, {Elsworth}, {Gruberbauer}, {Guenther}, {Stello}, {Basu},
  {Garc{\'{\i}}a}, {Chaplin}, {Mullally}, {Still}, \&
  {Thompson}}]{kallinger2012}
---, 2012, A\&A, 541, A51

\bibitem[{{Kjeldsen} \& {Bedding}(1995)}]{kb1995}
{Kjeldsen} H., {Bedding} T.~R., 1995, \aap, 293

\bibitem[{{Montalb{\'a}n} {et~al}\mbox{.}(2010){Montalb{\'a}n}, {Miglio},
  {Noels}, {Scuflaire}, \& {Ventura}}]{montalban2010}
{Montalb{\'a}n} J., {Miglio} A., {Noels} A., {Scuflaire} R., {Ventura} P.,
  2010, ApJ, 721, L182

\bibitem[{{Mosser} \& {Appourchaux}(2009)}]{mosser2009}
{Mosser} B., {Appourchaux} T., 2009, \aap, 508, 877

\bibitem[{{Mosser} {et~al}\mbox{.}(2011{\natexlab{a}}){Mosser}, {Barban},
  {Montalb{\'a}n}, {Beck}, {Miglio}, {Belkacem}, {Goupil}, {Hekker}, {De
  Ridder}, {Dupret}, {Elsworth}, {Noels}, {Baudin}, {Michel}, {Samadi},
  {Auvergne}, {Baglin}, \& {Catala}}]{mosser2011mm}
{Mosser} B. {et~al.}, 2011{\natexlab{a}}, A\&A, 532, A86

\bibitem[{{Mosser} {et~al}\mbox{.}(2011{\natexlab{b}}){Mosser}, {Belkacem},
  {Goupil}, {Michel}, {Elsworth}, {Barban}, {Kallinger}, {Hekker}, {De Ridder},
  {Samadi}, {Baudin}, {Pinheiro}, {Auvergne}, {Baglin}, \&
  {Catala}}]{mosser2011up}
---, 2011{\natexlab{b}}, A\&A, 525, L9

\bibitem[{{Mosser} {et~al}\mbox{.}(2014){Mosser}, {Benomar}, {Belkacem},
  {Goupil}, {Lagarde}, {Michel}, {Lebreton}, {Stello}, {Vrard}, {Barban},
  {Bedding}, {Deheuvels}, {Chaplin}, {De Ridder}, {Elsworth}, {Montalban},
  {Noels}, {Ouazzani}, {Samadi}, {White}, \& {Kjeldsen}}]{mosser2014}
---, 2014, \aap, 572, L5

\bibitem[{{Mosser} {et~al}\mbox{.}(2012{\natexlab{a}}){Mosser}, {Elsworth},
  {Hekker}, {Huber}, {Kallinger}, {Mathur}, {Belkacem}, {Goupil}, {Samadi},
  {Barban}, {Bedding}, {Chaplin}, {Garc{\'{\i}}a}, {Stello}, {De Ridder},
  {Middour}, {Morris}, \& {Quintana}}]{mosser2012}
---, 2012{\natexlab{a}}, \aap, 537, A30

\bibitem[{{Mosser} {et~al}\mbox{.}(2012{\natexlab{b}}){Mosser}, {Goupil},
  {Belkacem}, {Marques}, {Beck}, {Bloemen}, {De Ridder}, {Barban}, {Deheuvels},
  {Elsworth}, {Hekker}, {Kallinger}, {Ouazzani}, {Pinsonneault}, {Samadi},
  {Stello}, {Garc{\'{\i}}a}, {Klaus}, {Li}, {Mathur}, \&
  {Morris}}]{Mosser2012rot}
---, 2012{\natexlab{b}}, \aap, 548, A10

\bibitem[{{Mosser} {et~al}\mbox{.}(2015){Mosser}, {Vrard}, {Belkacem},
  {Deheuvels}, \& {Goupil}}]{mosser2015}
{Mosser} B., {Vrard} M., {Belkacem} K., {Deheuvels} S., {Goupil} M.~J., 2015,
  \aap, 584, A50

\bibitem[{{Pietrinferni} {et~al}\mbox{.}(2004){Pietrinferni}, {Cassisi},
  {Salaris}, \& {Castelli}}]{pietrinferni2004}
{Pietrinferni} A., {Cassisi} S., {Salaris} M., {Castelli} F., 2004, \apj, 612,
  168

\bibitem[{{Pinsonneault} {et~al}\mbox{.}(2014){Pinsonneault}, {Elsworth},
  {Epstein}, {Hekker}, {M{\'e}sz{\'a}ros}, {Chaplin}, {Johnson},
  {Garc{\'{\i}}a}, {Holtzman}, {Mathur}, {Garc{\'{\i}}a P{\'e}rez}, {Silva
  Aguirre}, {Girardi}, {Basu}, {Shetrone}, {Stello}, {Allende Prieto}, {An},
  {Beck}, {Beers}, {Bizyaev}, {Bloemen}, {Bovy}, {Cunha}, {De Ridder},
  {Frinchaboy}, {Garc{\'{\i}}a-Hern{\'a}ndez}, {Gilliland}, {Harding},
  {Hearty}, {Huber}, {Ivans}, {Kallinger}, {Majewski}, {Metcalfe}, {Miglio},
  {Mosser}, {Muna}, {Nidever}, {Schneider}, {Serenelli}, {Smith}, {Tayar},
  {Zamora}, \& {Zasowski}}]{pinsonneault2014}
{Pinsonneault} M.~H. {et~al.}, 2014, \apjs, 215, 19

\bibitem[{Schwarz(1978)}]{schwarz1978}
Schwarz G., 1978, Ann. Statist., 6, 461

\bibitem[{{Stello} {et~al}\mbox{.}(2016){Stello}, {Cantiello}, {Fuller},
  {Huber}, {Garc{\'{\i}}a}, {Bedding}, {Bildsten}, \& {Silva
  Aguirre}}]{stello2016}
{Stello} D., {Cantiello} M., {Fuller} J., {Huber} D., {Garc{\'{\i}}a} R.~A.,
  {Bedding} T.~R., {Bildsten} L., {Silva Aguirre} V., 2016, \nat, 529, 364

\bibitem[{{Stello} {et~al}\mbox{.}(2013){Stello}, {Huber}, {Bedding},
  {Benomar}, {Bildsten}, {Elsworth}, {Gilliland}, {Mosser}, {Paxton}, \&
  {White}}]{stello2013}
{Stello} D. {et~al.}, 2013, ApJ, 765, L41

\bibitem[{{Tassoul}(1980)}]{tassoul1980}
{Tassoul} M., 1980, \apjs, 43, 469

\bibitem[{{Vrard}, {Mosser} \& {Samadi}(2016){Vrard}, {Mosser}, \&
  {Samadi}}]{vrard2016}
{Vrard} M., {Mosser} B., {Samadi} R., 2016, \aap, 588, A87

\bibitem[{{Yu} {et~al}\mbox{.}(2016){Yu}, {Huber}, {Bedding}, {Stello},
  {Murphy}, {Xiang}, {Bi}, \& {Li}}]{yu2016}
{Yu} J., {Huber} D., {Bedding} T.~R., {Stello} D., {Murphy} S.~J., {Xiang} M.,
  {Bi} S., {Li} T., 2016, \mnras, 463, 1297

\end{thebibliography}
